\documentclass{aa}
\usepackage{graphicx}
\usepackage{graphics}
\usepackage{txfonts}
\def\clus{IC\,2391}
\def\A{{\AA}}

\def\i{{\sc{i}}}
\def\ii{{\sc{ii}}}
\def\abn{$[{N_{\mathrm{El}}}/{N_{\mathrm{tot}}}]$}
\def\Teff{$T_{\mathrm{eff}}$}
\def\logg{\ensuremath{\log g}}
\def\vmic{$v_{\mathrm{mic}}$}

\def\vsi{\ensuremath{v\sin i}}
\def\kms{$\mathrm{kms}^{-1}$}
\def\exc{$\chi_{\mathrm{excit}}$}
\def\gf{log$gf$}
\def\web#1{{\sc{#1}}}
\begin{document}
   \title{Abundance analysis of 5 early-type stars in the young open cluster 
          \clus\thanks{Based on observations made with ESO Telescopes at the Paranal
                       Observatory under programme ID 266.D-5655}}
    \authorrunning{Ch. St\"utz et al.}

   \author{Ch. St\"utz\inst{1,2},
          S. Bagnulo\inst{2}, E. Jehin\inst{2}, 
          C. Ledoux\inst{2}, R. Cabanac\inst{3}, C. Melo\inst{2}
          \and
          J.V. Smoker\inst{2,4}
          }

   \offprints{Ch. St\"utz}

   \institute{Institute of Astronomy (IfA), University of Vienna,
              T\"urkenschanzstrasse 17, A-1180 Vienna, Austria
         \and
              European Southern Observatory, 
              Casilla 19001, Santiago 19, Chile
         \and
              Canada-France-Hawaii Telescope Corporation,
			  65-1238 Mamalahoa Hwy. Kamuela, Hawaii 96743, USA
         \and Astrophyisics \& Planetary Science Research Division, 
              Department of Physics and Astronomy, The Queen's University 
              of Belfast, University Road, Belfast, BT7 1NN, U.K.
             }

   \date{Received; accepted}

   \abstract {} 
{It is unclear whether chemically peculiar stars of the upper main
sequence represent a class completely distinct from normal A-type
stars, or whether there exists a continuous transition from the normal
to the most peculiar late F- to early B-type stars. A systematic
abundance analysis of open cluster early-type stars would help to
relate the observed differences of the chemical abundances of the
photospheres to other stellar characteristics, without being concerned
by possible different original chemical composition. Furthermore, if a
continuous transition region from the very peculiar to the so called
normal A--F stars exists, it should be possible to detect objects with
mild peculiarities.}
{As a first step of a larger project, an abundance analysis of 5 F--A
type stars in the young cluster \clus\ was performed using high
resolution spectra obtained with the UVES instrument of the ESO VLT.}
{Our targets seem to follow a general abundance pattern: close to
solar abundance of the light elements and iron peak elements, heavy
elements are slightly overabundant with respect to the sun, similar to
what was found in previous studies of normal field A-type stars of the
galactic plane.  We detected a weakly chemically peculiar star,
HD\,74044. Its element pattern contains characteristics of CP1 as well
as CP2 stars, enhanced abundances of iron peak elements and also
higher abundances of Sc, Y, Ba and Ce.
We did not detect a magnetic field in this star (detection limit was 2\,kG).
We also studied the star SHJM\,2, proposed as a pre-main sequence object in
previous works. Using spectroscopy we found a high surface gravity, which 
suggests that the star is very close to the ZAMS.}
{}
   \keywords{
              Open clusters and associations: individual: IC2391 --
              Stars: abundances -- 
              Stars: evolution --
              Stars: chemically peculiar --
              Stars: pre-main sequence
            }

   \maketitle
%

%
\section{Introduction}
The atmospheres of main sequence A-type stars should be relatively
simple to understand and to model. Convection is apparently absent,
stellar winds are very weak, and little or no photospheric
microturbulence is generally present.  Yet, in this group of
stars, a large variety of peculiarities are observed.  Among A- and
B-type stars, about 10\,\% are chemically peculiar, i.e., the
analysis of their spectra reveal overabundances (e.g., of iron-peak
and/or rare-earths elements) and/or underabundances (e.g., of He,
Ca, and Sc), compared to the composition of the solar
photosphere. Many of these chemically peculiar (CP) stars exhibit
large scale magnetic fields with a typical strength of a few hundreds up
to a few tens of thousands of Gauss. Furthermore, many of the CP stars
reveal very inhomogeneous atmospheric distributions of numerous
elements. Another characteristic of CP stars is that,
compared to the normal A and B-type stars, most have long rotation
periods -- typically several days, but up to a few decades for some
magnetic CP stars. For a more detailed introduction to the variety of
phenomena observed in A and B-type stars see, e.g., Preston
(\cite{Preston74}), Wolff (\cite{Wolff83}).

An important question to address is whether the CP stars represent a
group that is well separated from ``normal'' A and B-type stars, or
whether CP stars are the extreme cases of a group of stars
spanning all grades of peculiarity.  If a continuous transition from
the very peculiar to the so called normal A--F stars exists, it should
be possible to detect objects with mild peculiarities.

There are several reasons why these stars are so difficult to
identify. Rotation velocities of normal A-type stars are in most cases
too high ($\langle \vsi \rangle \sim$ 150\,\kms) to allow precise
abundance determination. Pronounced chemical peculiarity is mostly
found in slow rotators ($\langle \vsi \rangle \sim$ 40\,\kms), whereas
studying a transition from CP to normal stars also implies analysing
objects with higher projected rotation velocities. Thus it is not
clear whether this anticorrelation of \vsi\ and peculiarity is a
physical one, or if it is due to a selection effect.  In other words,
is rotation preventing chemical peculiarity or is it hiding chemical
peculiarity?  Hill and Landstreet (\cite{HillLandstreet}) explicitly
showed for narrow lined A stars ($\vsi \la 25$\,\kms) of the galactic
plane that the variations in the individual elemental abundances can
be quite large ($\pm$\,0.4\,dex). The same is reflected
in the work of Adelman and collaborators (e.g. Kocer et
al. \cite{Kocer03}, Pintado \& Adelman \cite{Pintado03}).

The proper way to address the questions above is to perform a detailed 
abundance analysis of early--type stars of different ages 
and rotational velocities. Members of open clusters are of special 
interest because one can safely assume that members of the same cluster 
are more or less of the same age and chemical composition. They should differ
from each other only by their initial mass. Furthermore, the cluster age
can be determined with much more accuracy than the age of individual stars 
in the field. 
We thus carried out a detailed study of 5 main sequence stars of 
spectral type B -- F belonging to the young open cluster \clus. 
For our investigations we used spectra obtained with UVES, the high 
resolution spectrograph of the ESO VLT.
As the stars of \clus~we analysed are roughly the same age and were formed 
from similar initial chemical composition, we expect that if significant
deviations from the abundance pattern typical for \clus~are found, they 
will be nearly independent of these two stellar properties. Thus the 
probability that weak chemical peculiarity dates back to the birth of 
the star is higher than for stars located in the galactic field.
Since the cluster is very young ($\sim$ 36\,Mys, Lyng{\aa} \cite{Lynga1987}), 
main sequence (MS) evolutionary effects are small. One star of our sample, 
SHMJ\,2, may have not reached the MS yet.
\section{Observations and data reduction}
\subsection{Observations}
\clus\ was observed from 7 to 12 February 2001 with the UVES
instrument of the ESO VLT Unit 2 Kueyen within the framework of the
UVES Paranal Observatory Project (Bagnulo et
al. \cite{Bagnulo}). Target selection was performed mainly with the
WEBDA open cluster database
(\web{http://www.univie.ac.at/webda/}) developed by J-C.~Mermilliod at
the Institute for Astronomy of the University of Lausanne and
maintained by E. Paunzen at the University of Vienna. Some additional
pre-main sequence stars were selected using the work by Stauffer et
al. (\cite{Stauffer89}). Spectra for about 50 candidate cluster
members were obtained using the settings DIC1 (346+580) and DIC2
(437+860) (see UVES user manual VLT-MAN-ESO-13200-1825) with a 0.5''
slit width. The resulting spectra cover almost the entire spectral
range from 305\,nm to 1040\,nm with a spectral resolution of about
80\,000.  The raw data are available at the ESO archive under
programme ID 266.D-5655.
\subsection{Target selection}
In total, 50 stars of \clus\ were observed. For 8 early-type stars
among them, the rotational velocity is low enough and the quality of
the spectral data sufficient for our purposes of detailed abundance
analysis. These stars are listed in Table~\ref{sample}.

Information about membership was extracted from Robichon et al.
(\cite{Robichon99}), Levato et al. (\cite{Levato88}), Perry \& Hill
(\cite{PerryHill69}), and Stauffer et al. (\cite{Stauffer89}).
Membership was cross-checked using Hipparcos parallaxes, the most
recent $E(b-y)$ data found in the SIMBAD database and new radial
velocity measurements of UVES POP stars by Noterdaeme et al. (in
preparation).  For two objects of Table~\ref{sample}, membership
remains questionable.  HD\,75029 is not stated as a member in Perry \&
Hill (\cite{PerryHill69}), however, its astrometric parameters agree
well with the cluster mean. On the other hand HD\,73778, 
according to Levato et al. (\cite{Levato88}), very likely is not a 
member of \clus. The Si star HD\,74535 was investigated by 
L\"uftinger et al. (in prep.) parallel to our study.
%
\begin{table}[h]
\caption{Candidate target stars for abundance analysis. 
         Rotation velocities from Noterdaeme et al. (in preparation). 
         M? denotes questionable membership, \vsi~in \kms. The $\Delta a$ 
         index that is sensitive to CP2 type peculiarity
         is taken from Maitzen \& Catalano (\cite{MaitzenCatalano86}).
         A * indicates stars that were selected for the abundance analysis.
         } 
\label{sample}                    
\centering                        
\begin{tabular}{ l l c c l }      
\hline\hline                      
star           & sp.type  & m$_V$ & \vsi & notes       \\ 
\hline                                                    
HD\,73722      & F5\,V    & 8.92 & 6.8 &                    * \\  
HD\,73778      & F0\,V	  & 8.76 &  40 & M?                   \\
HD\,74044      & A3       & 8.48 &  32 & $\Delta a=1$ , Am? * \\
HD\,74275      & A0\,V	  & 7.27 &  62 &                    * \\
HD\,74535      & B8       & 5.55 &  37 & $\Delta a=26$ , Si   \\
HD\,75029      & A2/3	  & 9.45 &  24 & $\Delta a=-7$ , M? * \\
CPD\,-52\,1568 & F5\,V	  & 9.62 &  19 &                      \\
SHJM\,2	       & F9	      & 10.3 &  10 & PMS?               * \\
\hline                            
\end{tabular}
\end{table} 
%
We analysed HD\,73722, HD\,74044, HD\,74275,
HD\,75029 and SHJM\,2. This target selection is the result of a
compromise between keeping the spread of spectral types large and
the projected rotation velocities low.
%
\subsection{Data reduction}
Data have been reduced by the UVES POP team using an automatic
procedure based on the MIDAS UVES pipeline (Ballester et al.\
\cite{uvespipe}). Science frames were bias-subtracted and divided by
the extracted flat-field, except for the 860 setting, where
the 2D (pixel-to-pixel) flat-fielding was used in order to better
correct for fringing. Because of the high flux of the spectra
the \textit{average extraction} method was used instead of the
\textit{optimal extraction} method that is recommended for
spectra characterised by a signal to noise ratio (SNR) $\la
100$. Further details about the reduction procedure can be found in
Bagnulo et al.~(\cite{Bagnulo}).  All reduced spectra can be
obtained with the UVES POP web interface at
\web{http://www.eso.org/uvespop/}.

Taking advantage of the quality control parameters produced by the
UVES pipeline, we performed a check of the instrument
stability and actual performance. The accuracy of the wavelength
calibration was found to be of the order of 300\,ms$^{-1}$ only because 
the Th-Ar reference frames were taken the following morning and not
before and after each observation (the temperature difference between
science exposures and calibrations was typically 1\,K). For
this work we used the following wavelength intervals:
3730--4990\,\AA\ (characterised by a mean resolution 
$\langle R \rangle = 76\,000$ and a SNR of 100--150); 
4760--5770\,\AA\ ($\langle R \rangle = 82\,000$ and SNR = 210--150);
5840--6840\,\AA\ ($\langle R \rangle = 73\,000$ and SNR = 170--130).
We found a maximum variation of 3.7\,\% for these mean resolutions.
\subsection{Continuum normalisation}
The targets selected for the abundance analysis
include objects with rotation velocities up to $\simeq 60$\,\kms.
For such fast rotating stars, the determination of the elemental
abundances depends critically on the accuracy of the continuum
normalisation. To be sure to perform accurate continuum fitting,
we performed some tests using the spectra of HD~74169, a slow
rotating cluster member Ap star that will be analysed in detail in a
forthcoming paper by L\"uftinger et al. We normalised the star's high
SNR ($\approx$\,300) spectra in two independent ways:
Using the merged output of the UVES pipeline, 
and using the intermediate ``2D unmerged'' spectra, i.e.,
the spectra corresponding to the individual echelle orders (also
available through the UVES POP web interface). The fully merged
spectra are affected by some artefacts mainly due to an imperfect
merging of the echelle orders (for details see Bagnulo et
al. \cite{Bagnulo} and references therein), hence better results are
expected to be obtained by normalisation of the unmerged spectra.
However, an optimum normalisation could be obtained by dividing the 
merged spectra into sub-spectra of a maximum length of
100\,\A, and by treating these individually.  This way we obtained an
agreement to better than 1\,\% over the whole spectral range for the
two different techniques. The maximum size of the differences 
in the normalised spectra obtained with the two methods does not generally
occur at the wavelengths of the the extremes of the echelle
orders. This suggests that the order merging performed by the UVES
pipeline is sufficiently accurate and does not lead to major artefacts
in the abundance analysis.

Our comparison showed that we could not clearly define the continuum for
the hydrogen lines (which are in general broader than 100\,\A) and thus
we did not use them to determine fundamental atmospheric parameters.

To identify telluric lines we used the star HD\,74196 as
reference. The spectrum of this cluster member does not contain many
lines (spectral type B7V) which are heavily broadened due
to the high rotational velocity of \vsi~= 300\,\kms.
\section{Abundance analysis}
\begin{figure}[h]
   \centering
   \includegraphics[angle=-90,width=85mm]{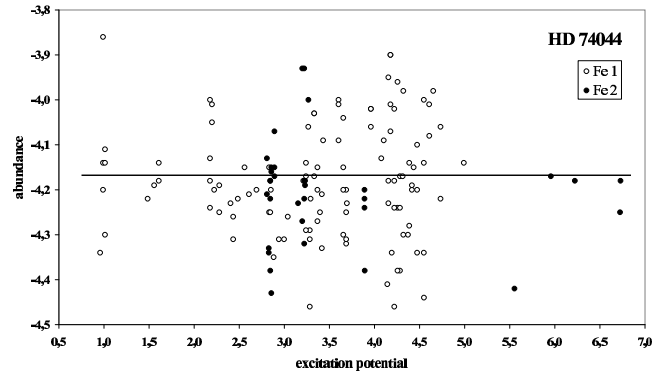}
      \caption{Excitation potential vs. abundance for the final values of
	           \Teff~and \logg~of HD\,74044. 
               Open circles -- Fe{\scriptsize{I}}, filled circles --
               Fe{\scriptsize{II}}. A least squares linear regression 
               yields: abn = -0.0002\exc~- 4.18. }
         \label{ABNvsX}
   \end{figure}

The starting values for the fundamental parameters of the atmosphere
of the selected stars were estimated via Str\"omgren
photometry (Johnson $UBV$ photometry in the case of SHJM\,2) and
evolutionary tracks for the cluster interpolated from Schaller et
al. (\cite{Schaller04}). The uncertainties of these estimates are
quite large (typically 200\,K in \Teff\ and 0.25 in \logg), thus we
tried to improve the fundamental parameters by spectroscopic means.
\Teff\ estimates were improved via elimination of an abundance\,--\,\exc\
correlation (see Fig.\,\ref{ABNvsX}).
\logg\ estimates were improved by making sure that there is no systematic
difference in abundances between different ionisation stages of 
chemical elements (called in the following "ionisation equation condition").
The presence of a magnetic field was checked looking at the correlation of
abundances with Land\'{e}factors, and by searching for magnetically
split lines.
%
%
Together with the mean metallicity of the cluster ([Fe/H]=-0.03,
Randich et al. \cite{Randich01}), these parameters defined the first
atmosphere calculated with LLmodels (Shuliak et al.
\cite{Shuliak04}) in ODF mode (ODF: estimating line absorption via a
pre calculated opacity distribution function).  Using this model we
calculated the emerging flux at the center of lines extracted from the
VALD database (Piskunov et al. \cite{VALD1}, Kupka et
al. \cite{VALD2}, Ryabchikova et al.  \cite{VALD3}) and preselected
only those that make an important contribution to the opacity 
($\kappa_{\nu\mathrm{L}}$/$\kappa_{\nu\mathrm{C}}$ $\ge$ 1\%, 
$\kappa_{\nu\mathrm{L}}$ and $\kappa_{\nu\mathrm{C}}$ are the line and 
continuum absorption coefficient). 

We determined the initial microturbulence and element abundance
pattern from equivalent widths of unblended lines using the {\em
widthV} software written by V.\ Tsymbal. Assuming that for our set of
spectral lines non--LTE and stratification effects are negligible,
there should be no correlation between abundances determined from the
equivalent widths and the widths themselves or the excitation energy, 
and the ionisation equation condition should be fulfilled. 
Eliminating the above correlations for the elements where we found 
the most unblended or marginally blended lines in the spectra  
(typically Fe, Cr, Ti, Ni, Ca) 
resulted not only in a fairly good determination of
microturbulence and first abundances, but also in an improvement of
effective temperature and surface gravity. For HD\,74275 we had to
perform this step via synthetic line fitting, since this star rotates
with $\vsi= 60.5$\,\kms\ and shows only a few unblended lines.
%
\begin{table*}[h]
\begin{minipage}[t]{\columnwidth}
\renewcommand{\footnoterule}{}  
\caption{Starting and final set of atmospheric parameters.
         UP -- UVES\,POP number} 
\label{atmospheres}             
\centering                      
\begin{tabular}{|ll|rrrr|rrrr|} 
\hline\hline                    
 & & \multicolumn{4}{c|}{initial set} & \multicolumn{4}{c|}{final set}  \\
\hline
star      & UP & \Teff\footnote{Obtained from Str\"omgren photometry}
		       & \logg$^{a}$  
		       & \logg\footnote{Obtained using tracks from Schaller \cite{Schaller04}}
               & \vsi\footnote{Taken from Noterdaeme et al. (in prep.)}
               & \Teff~[K]    & \logg~[cgs] & \vmic~[\kms] & \vsi \\
\hline                          
HD\,73722 & 04 &  6550 & 4.25 & 4.35 & 6.8  &  6480 & 4.30 & 1.30 &  6.7 \\
HD\,74044 & 11 &  8000 & 4.20 & 4.30 & 32   &  8130 & 4.45 & 2.65 & 33.7 \\
HD\,74275 & 23 & 10500 & 4.42 & 4.29 & 62   & 10200 & 4.40 & 1.30 & 60.5 \\
HD\,75029 & 43 &  7770 & 4.20 & 4.30 & 24   &  7800 & 4.30 & 2.00 & 22.0 \\
SHJM\,2 \footnote{Taken from Stauffer et.al (\cite{Stauffer89})}
          & 73 &  5950\footnote{Obtained from Johnson UBV photometry} 
                       &      & 4.40 & 10.3 &  6100 & 4.40 & 1.15 & 10.6 \\
\hline                                   
\end{tabular}
\end{minipage}
\end{table*}
%
For synthetic line fitting all our models were calculated with LLmodels in the
line-by-line modus (Shuliak \cite{Shuliak04}). This means that the individual 
abundance pattern is included in our atmospheric models as well. The hydrogen
lines were treated using VCS theory (Vidal, Cooper and Smith \cite{VCS73}).
Convection was modelled according to the formalism of Canuto and 
Mazzitelli (\cite{CM91}). For atmospheric modelling as well as for the line 
synthesis with Synth3 (Piskunov \cite{Piskunov92} and 
Valenti et al. \cite{Valenti98}) we extracted the atomic 
lines from the VALD line database. To analyse the spectra, in the sense of
deriving noise, deviations between observed and synthesized spectra,
measuring equivalent widths, etc., we developed the 'Little Spectrum 
Analyser' (Lispan). 
The graphical interface for the codes which also controls the automatic
line core fitting is called ATC.
The set of atmospheric modelling and analysis software used for this 
investigation can be reviewed on the web at 
\web{http://ams.astro.univie.ac.at/computer/ChrSoft/chrsoft.html}\\
Once we had fixed \vsi~and \vmic~(see Table~\ref{atmospheres}) by synthetic 
line fitting of the unblended lines, we determined the elemental abundances in 
an iterative process:
\begin{list}{$\bullet$}{}
 \item Automatic line core fitting with ATC, Synth3 and Lispan.
       This is faster than fitting the whole line, but one has to be 
       sure of \vsi, \vmic~and if present, the magnetic field. 
	   The only free parameters here are the abundances.
 \item Check for peculiarities and magnetic fields.
 \item Fine tuning of fundamental parameters (\Teff, \logg, \vmic, \vsi). 
       Check the conditions mentioned above and whether they hold when 
       peculiarities have been found.
 \item Edit line selection if blends have been overlooked, certain lines 
       show non--LTE effects, the quality of line parameters are uncertain,
       etc. 
\end{list}
This procedure resulted in fundamental atmospheric parameters with internal
errors lower than 150\,K in \Teff, 0.20\,dex in \logg, 0.20\,\kms~in \vmic, 
0.3\,\kms~in \vsi~ and accurate element abundances (0.1 -- 0.3\,dex 
in general) for all the program stars except HD\,75029 which very likely is
a spectroscopic binary.
For more detailed discussions on internal and absolute errors in modern 
abundance analyses see Andrievsky et al.\ (\cite{Andrievsky02}) or 
Hill \& Landstreet (\cite{HillLandstreet}).
\section{Results}
The atmospheric parameters of our stars were checked by fits of the H
lines. Atomic lines blended by the wings of H Balmer lines were
not used for abundance analysis because we could not precisely 
determine the continuum. 
In some cases the number of lines suitable for deriving
abundances of a certain element was very small. 
For four stars these lines are listed in Table~\ref{lines}.
For the star HD\,74044 the complete list of lines used for the 
abundance analysis is presented in Table~\ref{linesCP}. 
We found no indications for mean magnetic surface fields larger than 1\,kG
in HD\,73722, HD\,75029 and SHJM\,2, or larger than 2\,kG in HD\,74044 and 
HD\,74275.
We did not find any published high quality abundance analyses 
of stars of \clus~for a comparison with our work.
In the following we will present the results for the 5 stars we have 
analysed.

\begin{figure}[h] 
 \centering
 \includegraphics[angle=-90,width=85mm]{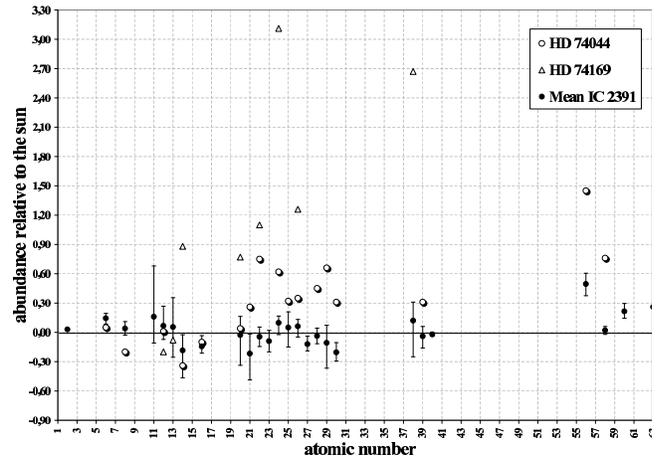}
    \caption{Comparison of the abundances in HD\,74044 (open circles) with
             those of the CP star HD\,74169 (open triangles, 
             L\"uftinger, in prep.) and to
             the mean of the other members of \clus~(filled circles).
             Abundances are in units of \abn~- \abn$_{\odot}$ (solar abundances
             according to Grevesse and Sauval \cite{GrevesseSauval}).
             Error bars indicate maximum deviations from the mean.
           }
       \label{nr11}
\end{figure}

\paragraph{HD\,74044:}
For HD\,74044 we determined a projected rotation velocity of
33.7\,\kms.  Fundamental atmospheric parameters were derived 
from Fe, Ca, Cr, Mg, Ni and Ti. We could determine
relatively precise abundances for 10 elements and abundance estimates
for C, Ce, Cu, Mn, O, Y and Zn (see Table~\ref{abundance}).
Comparing the abundance pattern of HD\,74044 to the other stars
we see indications for a mild chemical peculiarity, which is supported by
the higher abundances of Sc, Y, Ba and Ce (Fig.\,\ref{nr11}).
The element pattern of HD\,74044 contains characteristics of CP1 stars
(enhanced iron peak elements) as well as CP2 stars (enhanced Sc, and
heavy elements). A similar pattern but much more pronounced can be seen 
in the cluster star HD\,74169 known to be of type CP2 
(L\"uftinger et al, in prep.). 
Due to the rotation velocity of 33.7\,\kms our
detection limit for a mean surface magnetic field is about
2\,kG. Polarimetric observations will be necessary to
check for the presence a magnetic field.

\paragraph{HD\,73722:}
This slowly rotating star (\vsi~= 6.7\,\kms) is listed in the
catalogue of eclipsing and spectroscopic binary stars by
Popova \& Kraicheva (\cite{PK84}). But after carefully scanning
our spectra we found no hints of line patterns that could originate from
a stellar companion or a close background star.
Temperature, surface gravity and microturbulence (Table \ref{atmospheres}) 
were confirmed with lines from Fe, Cr, Ti, Ni, Ca, Si and Mn. Ti\i~and
Ti\ii~deviated noticeably from the ionisation equilibrium condition. 
Their abundances are -7.08 and -7.16 respectively. Typical for this 
analysis was 
$|$ \abn$_{\mathrm{I}}$ - \abn$_{\mathrm{II}}$ $|$ $<$ 0.05 
for different ionisation stages of the same element. 
The abundance of Sr is uncertain because both lines we analysed 
(Sr\ii~4161.792\,\A~and Sr\i~4607.327\,\A) are blended.
The result for oxygen should be considered as an upper limit.
%
   \begin{figure}[h]
   \centering
   \includegraphics[width=90mm]{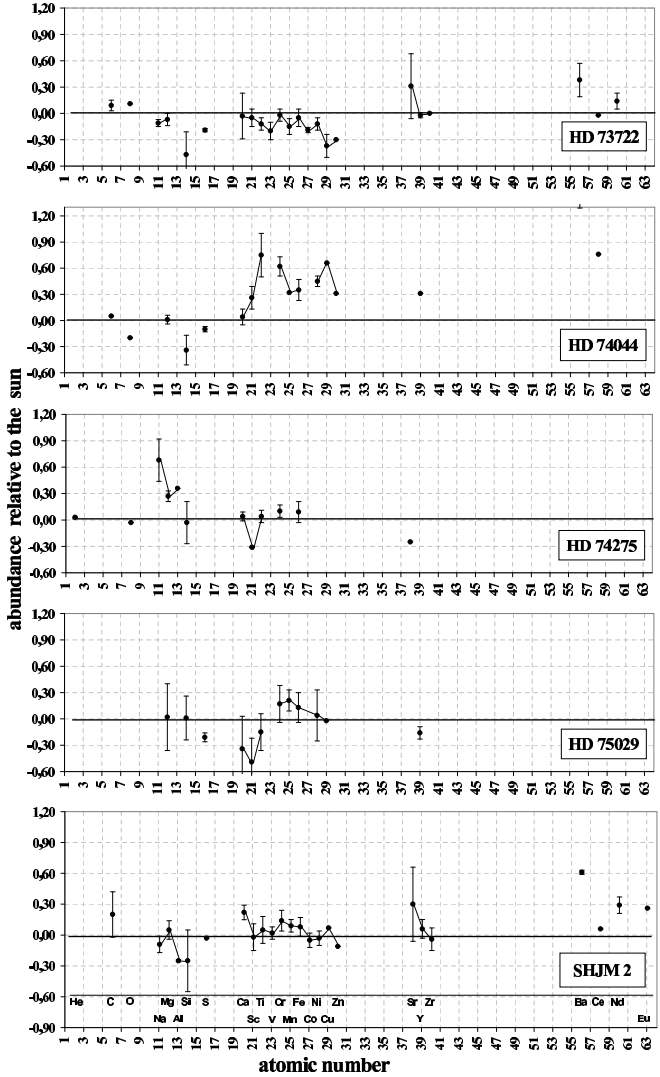}
      \caption{Elemental abundances relative to the sun.
               Solar values according to Grevesse and Sauval
               (\cite{GrevesseSauval}).
               Data points of elements with consecutive atomic numbers are
               connected with lines. Errors as in Table~\ref{abundance}.}
         \label{results}
   \end{figure}
%
\paragraph{HD\,74275:}
With $\vsi = 60.5$\,\kms, this is the fastest rotating star we have
analysed. It is also the hottest one. Thus only few lines could be
used and the quality of our Ca, O, Si and Na abundances is
not as good as our internal error estimates may suggest. The
atmospheric parameters were refined from Fe, Cr, Ti and Mg.
In 5 lines (Na\i~5889.951, Na\i~5895.924, Si\ii~6347.109,
Si\ii~6371.371 and He\i~5875.615) we corrected for atmospheric
features (see sect.2.4). The remarkable overabundance of Na
($\approx$0.7\,dex) is probably a non--LTE effect we could not account
for. Takeda (\cite{Takeda03}) and Asplund (\cite{Asplund05}) both
suggest non--LTE abundance corrections of up to -0.5\,dex for these two
lines.
The abundance of helium, determined from 1 line (He\i~5875.615\A), 
is more or less solar.

\paragraph{HD\,75029:}
Although the membership of this star may be questionable, we
included it in our analysis, since it was the only slowly rotating
star (\vsi~= 22\,\kms) with spectral type between A0 and F5 and no
known peculiarities. \textit{A posteriori} we found that its
abundances fit quite well in the overall pattern we see for \clus.
The star very likely has a faint fast rotating spectroscopic
companion, also noticed by Popova \& Kraicheva (\cite{PK84}).
We could not directly separate the spectrum of the companion, 
but we found our continuum to be systematically too low for regions 
around deeper lines. Strong contamination
due to twilight can be ruled out as the cause because we do not see this
behaviour in the spectra of HD\,74275 which has been observed at a
time even closer to dawn.  Unfortunately no details about the possible 
companion are known so far, hence our results for HD\,75029 are
not as accurate as they should be.

\paragraph{SHJM\,2:}
Using the UVES observations we could analyse in detail the
spectrum of a very slowly rotating (\vsi~= 10.6\,\kms) possible PMS
star.  Stauffer et al. (\cite{Stauffer89} and \cite{Stauffer97})
suggested that SHJM\,2 has not yet reached the ZAMS. The surface
gravity we determined by spectroscopic means ($\logg = 4.40$) is
surprisingly high if we consider the star to be a PMS object. Evolutionary
tracks for cluster stars give the same value for cluster members with
this effective temperature situated on the ZAMS.  Therefore SHJM\,2 is
probably at the very end of its pre main sequence phase or
has already reached the ZAMS.  Since SHJM\,2 is already in the final
stage of its PMS phase and very close to the main sequence, we applied
the techniques developed for main sequence stars for an abundance
analysis. Examining the iron lines Fe\ii~6149.258 and
Fe\i~6336.8243 we could detect no magnetic field.  The star generally
reflects the pattern of the other cluster members (except HD\,74044 of
course).
\section{Discussion and conclusions}
We have carried out a detailed analysis of the atmospheres of 5 A--F stars 
in the cluster \clus~and have obtained reasonably accurate fundamental 
parameters and element abundances accurate within 0.1 -- 0.3\,dex. 
Specific element patterns were included in our atmospheric models and 
the presence of magnetic fields and the possibility of stratification were
checked too (when \vsi~$\le$ 25\,\kms).
\begin{figure}
 \centering
 \includegraphics[width=85mm]{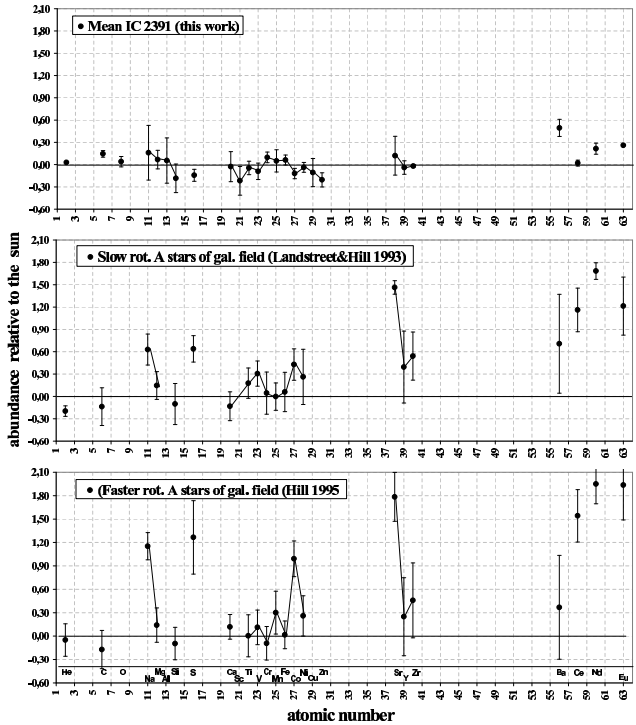}
    \caption{Mean abundance pattern of \clus~(HD\,74044 excluded)
             derived in this work compared to the abundance pattern found for
             field stars by Hill \& Landstreet (\cite{HillLandstreet}) and
             Hill (\cite{Hill95}). Elemental abundances are relative to the 
             sun with solar values according to Grevesse and Sauval
             (\cite{GrevesseSauval}).
             Data points of elements with consecutive atomic numbers are
             connected with lines. Error bars are variances of the mean
             abundances.}
     \label{compare}
\end{figure}

Understanding HD\,74044 contributes to our knowledge          
of the A star phenomenon. This star shows a mild peculiarity of the
type CP1 and CP2: Sc, Ti, Cr, Mn, Fe, Ni, Cu, Zn, Y and Ce are enhanced by 
0.3 -- 0.8\,dex relative to the mean of the other cluster members,
and even more so Ba.
Nevertheless we neither can confidently assign HD\,74044 to the 
group of CP2 nor CP1 stars. We situate this object in the
continuous transition region from the very peculiar to the normal
A--F stars.
We did not detect a magnetic surface field, but our detection limit
is rather high (2\,kG). Therefore follow-up polarimetric observations 
are necessary to clarify its status.

The abundance star to star scatter of the other cluster stars is,
as expected, small (0.1 -- 0.3\,dex) compared to what was
found for stars in the galactic plane (0.3 -- 0.5\,dex) by 
Hill \& Landstreet (\cite{HillLandstreet}) and Hill (\cite{Hill95}) 
for 6 and 9 objects respectively.
A comparison of a cluster abundance pattern derived in this work 
for 26 elements to the abundances of early A stars in the galactic plane 
is shown in Figure~\ref{compare}. Note that not only the star to star
variations of element abundances are larger in the galactic plane, but they 
also seem to increase for faster rotating stars. The latter may not be an 
intrinsic effect of rotation, but due to the fact that detailed abundance 
studies become rather difficult for faster rotating stars. 
For the needed accuracy our technique is limited to a 
\vsi~$\le$ 60.5\,\kms, measured for HD\,74275.

Our sample includes the star SHJM\,2, probably a PMS star 
already close to the ZAMS considering the findings of Stauffer et al.
(\cite{Stauffer89}) and the star's high surface gravity.
Assuming that its atmosphere can be modelled applying the same
physics and approximations as for MS A--F stars, we performed a detailed
abundance analysis in the optical range for this star. 
Its abundances reflect the general abundance pattern of \clus.
%
\begin{table*}[h]
\begin{minipage}[t]{\columnwidth}
\renewcommand{\footnoterule}{}    
\caption{Abundances \abn~of members of \clus. 
         Solar values according to Grevesse and Sauval
         (\cite{GrevesseSauval})} 
\label{abundance}                 
\centering                        
\begin{tabular}{ |l|rrr|rrr|rrr|rrr|rrr|r| }    
\hline\hline
    & \multicolumn{3}{c|}{{\bf{HD\,73722}}} 
	                 & \multicolumn{3}{c|}{{\bf{HD\,74044}}}
					                  &  \multicolumn{3}{c|}{{\bf{HD\,74275}}} 
									                   &  \multicolumn{3}{c|}{{\bf{HD\,75029}}}
													                    & \multicolumn{3}{c|}{{\bf{SHJM\,2}}} 
                                                                                          &{\bf{sun}}\\
El & abn\footnote{~abn~\hspace{4mm}element abundance:~
                              \abn~=~$^{10}log({N_\mathrm{Element}}/{N_\mathrm{Total}})$} 
           &(err)\footnote{~(err)\hspace{3mm}\,internal errors in units of the last digit}
                &[n]\footnote{~[n]\,\hspace{5mm}number of lines used for the final iteration step}
                     & abn &(err)&[n]& abn &(err)&[n]& abn &(err)&[n]& abn &(err)&[n] &abn\\
\hline                          
He&      &    &    &      &    &    &-1.08&(--)& [1]&     &    &    &      &    &     & -1.11\\
C & -3.43& (6)& [4]& -3.47&(--)& [2]&     &    &    &     &    &    & -3.32&(22)& [2] & -3.52\\
O & -3.10&(--)& [1]& -3.41&(--)& [2]&-3.24&(--)& [2]&     &    &    &      &    &     & -3.21\\
Na& -5.82& (4)& [4]&      &    &    &-5.03&(24)& [2]&     &    &    & -5.80& (8)& [4] & -5.71\\
Mg& -4.53& (7)& [5]& -4.45& (5)& [6]&-4.19& (6)& [7]&-4.44&(38)& [5]& -4.41& (9)& [3] & -4.46\\
Al&      &    &    &      &    &    &-5.21&(--)& [1]&     &    &    & -5.82& (1)& [2] & -5.57\\
Si& -4.96&(26)&[20]& -4.83&(17)& [6]&-4.52&(24)& [3]&-4.48&(25)&[10]& -4.74&(30)&[19] & -4.49\\
S & -4.90& (2)& [4]& -4.81& (3)& [3]&     &    &    &-4.92& (5)& [6]& -4.74&(--)& [1] & -4.71\\
  &      &    &    &      &    &    &     &    &    &     &    &    &      &    &     &      \\
Ca& -5.71&(26)&[29]& -5.64& (9)&[10]&-5.64& (5)& [2]&-6.02&(37)&[10]& -5.46& (7)&[13] & -5.68\\
Sc& -8.92&(10)& [9]& -8.61&(13)& [3]&-9.18&(--)& [1]&-9.36&(27)& [3]& -8.89&(13)& [6] & -8.87\\
Ti& -7.14& (7)&[51]& -6.27&(25)&[15]&-6.98& (7)& [6]&-7.17&(21)&[18]& -6.97&(13)&[29] & -7.02\\
V & -8.24&(10)& [5]&      &    &    &     &    &    &     &    &    & -8.02& (6)& [9] & -8.04\\
Cr& -6.39& (7)&[37]& -5.75&(11)&[21]&-6.27& (7)& [8]&-6.20&(21)&[24]& -6.23&(10)&[20] & -6.37\\
Mn& -6.80& (9)&[18]& -6.33&(--)& [1]&     &    &    &-6.44&(12)& [2]& -6.56& (6)& [9] & -6.65\\
Fe& -4.59&(10)&[298]&-4.19&(12)&[147]&-4.45&(12)&[29]&-4.41&(17)&[96]& -4.46& (9)&[169]& -4.54\\
Co& -7.31& (3)& [4]&      &    &    &     &    &    &     &    &    & -7.17& (7)& [6] & -7.12\\
Ni& -5.91& (7)&[50]& -5.34& (6)&[11]&     &    &    &-5.75&(29)&[17]& -5.82& (7)&[32] & -5.79\\
Cu& -8.20&(13)& [2]& -7.17&(--)& [1]&     &    &    &-7.85&(--)& [1]& -7.76&(--)& [1] & -7.83\\
Zn& -7.74&(--)& [1]& -7.13&(--)& [1]&     &    &    &     &    &    & -7.55&(--)& [1] & -7.44\\
  &      &    &    &      &    &    &     &    &    &     &    &    &      &    &     &      \\
Sr& -8.76&(37)& [2]&      &    &    &-9.32&(--)& [1]&     &    &    & -8.77&(36)& [3] & -9.07\\
Y & -9.82& (3)& [6]& -9.49&(--)& [1]&     &    &    &-9.96& (7)& [2]& -9.74& (9)& [4] & -9.80\\
Zr& -9.44&(--)& [1]&      &    &    &     &    &    &     &    &    & -9.48&(11)& [2] & -9.44\\
Ba& -9.53&(19)& [3]& -8.46&(16)& [3]&     &    &    &     &    &    & -9.30& (2)& [2] & -9.91\\
Ce&-10.48&(--)& [2]& -9.70&(--)& [1]&     &    &    &     &    &    &-10.40&(--)& [2] &-10.46\\
Nd&-10.40& (9)& [5]&      &    &    &     &    &    &     &    &    &-10.25& (8)& [2] &-10.54\\
Eu&      &    &    &      &    &    &     &    &    &     &    &    &-11.27&(--)& [1] &-11.53\\
\hline                                   
\end{tabular}
\end{minipage}
\end{table*}
%
\begin{acknowledgements}
We like to thank Theresa L\"uftinger for insight
in her analysis of HD\,74169 and Pasquier Noterdaeme for his 
most recent \vsi~data for this cluster. We also thank Martin Stift
for his helpful suggestions on presentation of the results.
Ch.~St\"utz acknowledges ESO DGDF for a three month studentship at ESO
Santiago/Vitacura and the FWF (project P17890).
\end{acknowledgements}
%

%
%
%
\Online
\begin{table*}[h]
\begin{minipage}[t]{\columnwidth}
\renewcommand{\footnoterule}{}  
\caption{Lines used for abundance determination of HD\,74044.} 
\label{linesCP}             
\centering                      
\begin{tabular}{|lrr|lrr|lrr|lrr|}   
\hline\hline                    
Identity & $\lambda$[\A] & \gf & Identity & $\lambda$[\A] & \gf & Identity & $\lambda$[\A] & \gf & Identity & $\lambda$[\A] & \gf \\
\hline                          
C1 & 4932.049 & -1.884 & Cr2 & 5308.408 & -1.846 & Fe1 & 5107.447 & -3.087 & Fe1 & 5572.842 & -0.275\\
C1 & 6014.834 & -1.585 & Cr2 & 5310.687 & -2.280 & Fe1 & 5107.641 & -2.418 & Fe1 & 5576.089 & -1.000\\
O1 & 6158.176 & -0.996 & Cr2 & 5313.563 & -1.650 & Fe1 & 5121.639 & -0.810 & Fe1 & 5586.756 & -0.120\\
O1 & 6158.186 & -0.409 & Cr2 & 5334.869 & -1.562 & Fe1 & 5125.117 & -0.140 & Fe1 & 5615.644 & 0.050\\
Mg2 & 4481.126 & 0.740 & Cr2 & 5407.604 & -2.151 & Fe1 & 5133.689 & 0.140 & Fe1 & 5624.542 & -0.755\\
Mg2 & 4481.150 & -0.560 & Cr2 & 5478.365 & -1.908 & Fe1 & 5139.252 & -0.741 & Fe1 & 5633.947 & -0.270\\
Mg1 & 4702.991 & -0.666 & Cr2 & 5508.606 & -2.110 & Fe1 & 5139.463 & -0.509 & Fe1 & 5686.530 & -0.446\\
Mg1 & 5172.684 & -0.402 & Cr2 & 5510.702 & -2.452 & Fe1 & 5148.036 & -0.629 & Fe1 & 5709.378 & -1.028\\
Mg1 & 5183.604 & -0.180 & Cr2 & 6089.632 & -1.265 & Fe1 & 5148.225 & -0.274 & Fe1 & 5731.762 & -1.300\\
Mg1 & 5528.405 & -0.620 & Mn1 & 6021.819 & 0.034 & Fe1 & 5159.058 & -0.820 & Fe1 & 5775.081 & -1.298\\
Si1 & 5675.417 & -1.030 & Fe1 & 4438.343 & -1.630 & Fe1 & 5162.273 & 0.020 & Fe1 & 5862.353 & -0.058\\
Si1 & 5747.667 & -0.780 & Fe1 & 4447.717 & -1.342 & Fe2 & 5197.577 & -2.100 & Fe1 & 5930.180 & -0.230\\
Si1 & 6145.016 & -0.820 & Fe1 & 4466.552 & -0.600 & Fe1 & 5202.336 & -1.838 & Fe1 & 6020.169 & -0.270\\
Si1 & 6243.815 & -0.770 & Fe2 & 4472.929 & -3.430 & Fe1 & 5229.845 & -1.127 & Fe1 & 6024.058 & -0.120\\
Si1 & 6244.466 & -0.690 & Fe1 & 4476.019 & -0.819 & Fe1 & 5229.880 & -0.236 & Fe1 & 6027.051 & -1.089\\
Si1 & 6254.188 & -0.600 & Fe1 & 4476.077 & -0.370 & Fe2 & 5234.625 & -2.230 & Fe1 & 6056.005 & -0.460\\
S1 & 6743.531 & -0.920 & Fe1 & 4484.220 & -0.864 & Fe1 & 5250.646 & -2.181 & Fe1 & 6065.482 & -1.530\\
S1 & 6748.837 & -0.600 & Fe1 & 4485.676 & -1.020 & Fe1 & 5253.462 & -1.573 & Fe2 & 6084.111 & -3.780\\
S1 & 6757.171 & -0.310 & Fe2 & 4491.405 & -2.700 & Fe2 & 5254.929 & -3.227 & Fe2 & 6113.322 & -4.110\\
Ca1 & 4425.437 & -0.286 & Fe1 & 4494.563 & -1.136 & Fe1 & 5273.164 & -0.993 & Fe1 & 6127.907 & -1.399\\
Ca2 & 5019.971 & -0.501 & Fe2 & 4508.288 & -2.250 & Fe1 & 5273.374 & -2.158 & Fe2 & 6147.741 & -2.721\\
Ca1 & 5581.965 & -0.569 & Fe2 & 4515.339 & -2.450 & Fe1 & 5281.790 & -0.834 & Fe2 & 6149.258 & -2.720\\
Ca1 & 5588.749 & 0.313 & Fe2 & 4520.224 & -2.600 & Fe1 & 5288.525 & -1.508 & Fe1 & 6191.558 & -1.417\\
Ca1 & 5590.114 & -0.596 & Fe2 & 4522.634 & -2.030 & Fe1 & 5302.302 & -0.720 & Fe1 & 6230.723 & -1.281\\
Ca1 & 5857.451 & 0.257 & Fe1 & 4528.614 & -0.822 & Fe2 & 5316.615 & -1.850 & Fe1 & 6232.641 & -1.223\\
Ca1 & 6122.217 & -0.386 & Fe2 & 4541.524 & -2.790 & Fe2 & 5316.784 & -2.760 & Fe2 & 6238.392 & -2.630\\
Ca1 & 6162.173 & -0.167 & Fe2 & 4555.893 & -2.160 & Fe1 & 5324.179 & -0.103 & Fe1 & 6252.555 & -1.687\\
Ca1 & 6439.075 & 0.394 & Fe2 & 4576.340 & -2.920 & Fe2 & 5325.553 & -3.120 & Fe1 & 6265.134 & -2.550\\
Ca1 & 6717.681 & -0.596 & Fe2 & 4583.837 & -1.860 & Fe1 & 5341.024 & -1.953 & Fe1 & 6336.824 & -0.856\\
Sc2 & 5031.021 & -0.400 & Fe1 & 4602.941 & -2.209 & Fe1 & 5353.374 & -0.840 & Fe2 & 6369.462 & -4.160\\
Sc2 & 5239.813 & -0.765 & Fe1 & 4607.647 & -1.545 & Fe2 & 5362.869 & -2.739 & Fe2 & 6383.722 & -2.070\\
Sc2 & 5526.790 & 0.024 & Fe1 & 4611.284 & -1.017 & Fe1 & 5367.467 & 0.443 & Fe1 & 6393.601 & -1.432\\
Ti2 & 4450.482 & -1.510 & Fe2 & 4620.521 & -3.240 & Fe1 & 5373.709 & -0.860 & Fe1 & 6411.649 & -0.595\\
Ti2 & 4468.507 & -0.600 & Fe1 & 4625.045 & -1.340 & Fe1 & 5383.369 & 0.645 & Fe2 & 6416.919 & -2.650\\
Ti2 & 4518.327 & -2.640 & Fe1 & 4632.912 & -2.913 & Fe1 & 5389.479 & -0.410 & Fe1 & 6419.950 & -0.240\\
Ti2 & 4529.474 & -1.650 & Fe2 & 4635.316 & -1.650 & Fe1 & 5391.461 & -0.825 & Fe1 & 6421.351 & -2.027\\
Ti2 & 4552.294 & -2.890 & Fe1 & 4643.463 & -1.147 & Fe1 & 5393.168 & -0.715 & Fe1 & 6430.846 & -2.006\\
Ti1 & 4552.453 & -0.340 & Fe1 & 4654.605 & -1.077 & Fe1 & 5400.502 & -0.160 & Fe2 & 6446.410 & -1.960\\
Ti2 & 4563.761 & -0.790 & Fe1 & 4678.846 & -0.833 & Fe1 & 5405.775 & -1.844 & Fe1 & 6677.987 & -1.418\\
Ti2 & 4571.968 & -0.230 & Fe1 & 4707.275 & -1.080 & Fe1 & 5410.910 & 0.398 & Ni1 & 4604.982 & -0.250\\
Ti2 & 4589.958 & -1.620 & Fe1 & 4903.310 & -0.926 & Fe1 & 5415.199 & 0.642 & Ni1 & 4714.408 & 0.260\\
Ti2 & 5010.212 & -1.300 & Fe1 & 4920.503 & 0.068 & Fe1 & 5424.068 & 0.520 & Ni1 & 4715.757 & -0.320\\
Ti2 & 5185.913 & -1.370 & Fe2 & 4923.927 & -1.320 & Fe2 & 5427.826 & -1.664 & Ni1 & 4904.407 & -0.170\\
Ti2 & 5188.680 & -1.050 & Fe1 & 4930.315 & -1.201 & Fe1 & 5429.505 & -1.016 & Ni1 & 5115.389 & -0.110\\
Ti2 & 5211.536 & -1.356 & Fe1 & 4946.388 & -1.170 & Fe1 & 5429.697 & -1.879 & Ni1 & 5176.559 & -0.440\\
Ti2 & 5418.751 & -2.110 & Fe1 & 4962.572 & -1.182 & Fe1 & 5429.827 & -0.527 & Ni1 & 5663.975 & -0.430\\
Ti2 & 5490.690 & -2.650 & Fe1 & 4966.089 & -0.871 & Fe2 & 5432.967 & -3.629 & Ni1 & 5694.977 & -0.610\\
Cr2 & 4539.595 & -2.290 & Fe1 & 4973.102 & -0.950 & Fe1 & 5434.524 & -2.122 & Ni1 & 5715.066 & -0.352\\
Cr2 & 4558.650 & -0.449 & Fe1 & 4962.572 & -1.182 & Fe1 & 5445.042 & -0.020 & Ni1 & 6086.276 & -0.530\\
Cr2 & 4565.740 & -1.820 & Fe1 & 4966.089 & -0.871 & Fe1 & 5446.917 & -1.914 & Ni1 & 6176.807 & -0.260\\
Cr2 & 4588.199 & -0.627 & Fe1 & 4973.102 & -0.950 & Fe1 & 5455.441 & 0.291 & Cu1 & 5105.537 & -1.516\\
Cr2 & 4634.070 & -0.990 & Fe1 & 4988.950 & -0.890 & Fe1 & 5455.610 & -2.091 & Zn1 & 4722.153 & -0.338\\
Cr1 & 4646.148 & -0.700 & Fe2 & 5004.195 & 0.497 & Fe1 & 5473.901 & -0.760 & Y2 & 5200.406 & -0.570\\
Cr1 & 4652.152 & -1.030 & Fe2 & 5018.440 & -1.220 & Fe1 & 5497.516 & -2.849 & Ba2 & 4554.029 & 0.170\\
Cr1 & 5204.506 & -0.208 & Fe1 & 5044.211 & -2.038 & Fe2 & 5529.932 & -1.875 & Ba2 & 5853.668 & -1.000\\
Cr2 & 5237.329 & -1.160 & Fe1 & 5049.820 & -1.355 & Fe1 & 5554.895 & -0.440 & Ba2 & 6141.713 & -0.076\\
Cr2 & 5249.437 & -2.489 & Fe2 & 5061.718 & 0.217 & Fe1 & 5560.212 & -1.190 & Ce2 & 4562.359 & 0.310\\
Cr2 & 5279.876 & -2.100 & Fe1 & 5065.014 & -0.134 & Fe1 & 5565.704 & -0.285 &  &  &\\
Cr2 & 5280.054 & -2.011 & Fe1 & 5074.748 & -0.200 & Fe1 & 5569.618 & -0.486 &  &  &\\
\hline                                   
\end{tabular}
\end{minipage}
\end{table*}
%
\begin{table*}[h]
\begin{minipage}[t]{\columnwidth}
\renewcommand{\footnoterule}{}  
\caption{Selected lines ([n] $<$ 5) used for abundance determination of
         HD\,73722, HD\,74275, HD\,75029 and SHJM\,2.} 
\label{lines}             
\centering                      
\begin{tabular}{|lrr|lrr|lrr|lrr|}   
\hline\hline                    
  \multicolumn{3}{c}{HD\,73722} & \multicolumn{3}{c}{HD\,74275} & \multicolumn{3}{c}{HD\,75029} & \multicolumn{3}{c}{SHJM\,2}\\
\hline
Identity & $\lambda$[\A] & \gf & Identity & $\lambda$[\A] & \gf & Identity & $\lambda$[\A] & \gf & Identity & $\lambda$[\A] & \gf \\
\hline                          
C1 & 4770.026 & -2.439 & He1 & 5875.615 & 0.409 & Sc2 & 5031.021 & -0.400 & C1 & 4932.049 & -1.884 \\
C1 & 4932.049 & -1.884 & O1 & 6156.756 & -0.899 & Sc2 & 5239.813 & -0.765 & C1 & 5380.337 & -1.842\\
C1 & 5052.167 & -1.648 & O1 & 6158.186 & -0.409 & Sc2 & 5526.790 & 0.024 & Na1 & 4497.657 & -1.560\\
C1 & 5380.337 & -1.842 & Na1 & 5889.951 & 0.117 & Mn1 & 5377.637 & -0.109 & Na1 & 5688.205 & -0.450\\
O1 & 6156.776 & -0.694 & Na1 & 5895.924 & -0.184 & Mn1 & 6021.819 & 0.034 & Na1 & 6154.226 & -1.560\\
Na1 & 5682.633 & -0.700 & Al2 & 4663.046 & -0.284 & Cu1 & 5105.537 & -1.516 & Na1 & 6160.747 & -1.260\\
Na1 & 5688.205 & -0.450 & Si2 & 5978.930 & 0.004 & Y2 & 5087.416 & -0.170 & Mg1 & 4167.271 & -1.004\\
Na1 & 6154.226 & -1.560 & Si2 & 6347.109 & 0.297 & Y2 & 5200.406 & -0.570 & Mg1 & 4571.096 & -5.691\\
Na1 & 6160.747 & -1.260 & Si2 & 6371.371 & -0.003 &  &  &  & Mg1 & 5711.088 & -1.833\\
S1 & 4695.443 & -1.920 & Ca1 & 4226.728 & 0.265 &  &  &  & Al1 & 6696.023 & -1.347\\
S1 & 4696.252 & -2.140 & Ca1 & 4454.779 & 0.335 &  &  &  & Al1 & 6698.673 & -1.647\\
S1 & 6052.583 & -1.330 & Sc2 & 4246.822 & 0.242 &  &  &  & S1 & 6052.674 & -0.740\\
S1 & 6757.171 & -0.310 & Sr2 & 4215.519 & -0.145 &  &  &  & Cu1 & 5105.537 & -1.516\\
Co1 & 4792.846 & -0.067 &  &  &  &  &  &  & Zn1 & 4722.153 & -0.338\\
Co1 & 5212.691 & -0.110 &  &  &  &  &  &  & Sr2 & 4161.792 & -0.502\\
Co1 & 5342.695 & 0.690 &  &  &  &  &  &  & Sr2 & 4215.519 & -0.145\\
Co1 & 5352.045 & 0.060 &  &  &  &  &  &  & Sr1 & 4607.327 & -0.570\\
Cu1 & 5105.537 & -1.516 &  &  &  &  &  &  & Y2 & 5087.416 & -0.170\\
Cu1 & 5218.197 & 0.476 &  &  &  &  &  &  & Y2 & 5119.112 & -1.360\\
Zn1 & 4722.153 & -0.338 &  &  &  &  &  &  & Y2 & 5402.774 & -0.510\\
Sr2 & 4161.792 & -0.502 &  &  &  &  &  &  & Y2 & 5544.611 & -1.090\\
Sr1 & 4607.327 & -0.570 &  &  &  &  &  &  & Zr2 & 4208.977 & -0.460\\
Zr2 & 4208.977 & -0.460 &  &  &  &  &  &  & Zr2 & 5112.297 & -0.590\\
Ba2 & 4166.000 & -0.420 &  &  &  &  &  &  & Ba2 & 4554.029 & 0.170\\
Ba2 & 5853.668 & -1.000 &  &  &  &  &  &  & Ba2 & 5853.668 & -1.000\\
Ba2 & 6496.897 & -0.377 &  &  &  &  &  &  & Ce2 & 4562.359 & 0.310\\
Ce2 & 4562.359 & 0.310 &  &  &  &  &  &  & Ce2 & 4628.161 & 0.220\\
Ce2 & 4628.161 & 0.220 &  &  &  &  &  &  & Nd2 & 4706.543 & -0.775\\
    &          &       &  &  &  &  &  &  & Nd2 & 5311.453 & -0.437\\
    &          &       &  &  &  &  &  &  & Eu2 & 4205.042 & 0.120\\
\hline                                   
\end{tabular}
\end{minipage}
\end{table*}
%
%
\end{document}